\documentclass[artice,aps,pra,showpacs,twocolumn,superscriptaddress]{revtex4}
\usepackage{graphicx,color}
\usepackage{amsmath,amsthm,amsfonts,amssymb,bm}
\usepackage{times}
\usepackage{epsf}
\usepackage[colorlinks={true}]{hyperref}
\usepackage[T1]{fontenc}
\usepackage[utf8]{inputenc}
\hypersetup{citecolor={blue}, filecolor={blue}, linkcolor={blue}, urlcolor={blue}}

\newcommand{\commentold}[1]{}
\DeclareMathSymbol{:}{\mathpunct}{operators}{"3A}
\usepackage[utf8]{inputenc}
\usepackage[english]{babel}
\usepackage{amsthm}
\theoremstyle{definition}

\usepackage{subfigure}
\begin{document}

\title{Quantum speed limit time for topological qubit influenced by fermionic and bosonic environment}

\author{Soroush Haseli}
\affiliation{Faculty of Physics, Urmia University of Technology, Urmia, Iran}
\email{soroush.haseli@uut.ac.ir}
\author{Hazhir Dolatkhah}
\affiliation{Department of Physics, University of Kurdistan, P.O.Box 66177-15175 , Sanandaj, Iran}
\author{Shahriar Salimi}
\affiliation{Department of Physics, University of Kurdistan, P.O.Box 66177-15175 , Sanandaj, Iran}


\date{\today}

\begin{abstract}{
Quantum theory sets a limit on the minimum time required to transform from an initial state to a target state. It is known as quantum speed limit time. quantum speed limit time can be used to determine the rate of quantum evolution for closed and open quantum systems. Given that in the real world we are dealing with open quantum systems, the study of quantum speed limit time for such systems has particular importance. In this work we consider the topological qubit realized by two Majorana modes. We consider the case in which the topological qubit is influenced by fermionic and bosonic environment. Fermionic and bosonic environments are assumed to have Ohmic-like spectral density. The quantum speed limit time is investigated for various environment with different Ohmic parameter. It is observed that  for super-Ohmic environment with increasing Ohmic parameter the quantum speed limit time gradually reaches to a constant value and thus the speed of evolution reaches to a uniform value. The effects of external magnetic field on the evolution rate are also studied. It is observed that with increasing magnitude of magnetic field, the quantum speed limit time decreases}
\end{abstract}
\maketitle
\section{Introduction}	
The minimum time required for the transformation of a quantum system from an initial state to a target state is known as quantum speed limit (QSL) time. It can be said that the QSL time stems from time-energy uncertainty principle. The maximum speed of quantum evolution can be obtained using QSL time. QSL time is used in many topics of quantum information theory such as quantum communication \cite{1}, investigation of
exact bounds in quantum metrology \cite{2}, computational bounds of physical systems \cite{3} and quantum optimal control algorithms \cite{4}. For closed quantum systems whose evolution is described using unitary operations, the QSL time is obtained using distance measures such as Bures angle and
relative purity \cite{5,6,7,8,9,10,11,12}. Among the most important QSL time bound for closed systems, one can mention two bounds, one is Mandelstam-Tamm (MT) bound \cite{11} and other is
Margolus-Levitin (ML) bound \cite{12}. Given that isolating a quantum system from its surroundings is difficult and almost impossible, and any real quantum system interacts inevitably with its surroundings, the study of open quantum systems is one of the fascinating topics in quantum information theory \cite{13,14,15}. Therefore, due to the importance of open quantum systems and their role in quantum information theory, the study of QSL time for these systems has been considered in many recent works \cite{16,17,18,19,20,21,22,23,24,25,26,27,28,29,30,31,32,33,34,35,36,37,38,39,40,41}.
In general, Mandelstam-Tamm bound and
Margolus-Levitin bound are used to describe QSL time in closed and open quantum systems. The generalizations of these two bounds for open quantum systems are given in Refs.\cite{7,8}. In Ref. \cite{24}, Deffner et al. present a comprehensive and unified bound for non-Markovian dynamics that includes both MT and ML bounds . In Ref. \cite{28}, Zhang et al. provide the QSL time bound  for arbitrary initial states .  They have shown that the QSL time is depend on quantum coherence of initial state.   Based on the definition they provide, QSL time is the minimum time required for the evolution of an open  quantum system from an initial state at time $\tau$ to target state at time $\tau+\tau_D$, where $\tau_D$ is driving time. In this work we will consider the QSL time bound which has introduced by  Zhang et al.\cite{28}.

It has been observed that topological quantum computing is a promising design for the realization of quantum computers with stable qubits \cite{Nayak}. According to recent studies, there exist different and new types of topological ordered states that are physically achievable, such as topological insulators and superconductors \cite{Fu,Hasan,Qi}. For these systems, some of the excitations are topologically protected, provided that some symmetries, such as time inversion, are maintained. In other words, the local perturbations that maintain these symmetries cannot
disentangle the topological excitations. The most interesting of these topological excitations are Majorana modes localized on topological defects, which follow the non-Abelian anyonic statistics \cite{Wilczek,Arovas,Ivanov}. The Kitaev 
1D spineless p-wave superconductor chain model is the most common model for realizing such Majorana modes.  Each on-site fermion can be decomposed into two Majorana modes. By properly adjusting the model, Majorana modes can be dangling at the end of the chain without pairing with other nearby Majorana  modes to form common fermions. So, these two separate Majorana modes can creat a topological qubit. There exist two meanings to the word topological here : One meaning is that it is composed of Majorana modes that are topological excitations, another implication is that the topological qubit itself is non-local, meaning that the two majorana states are very separate and therefore cannot be combined into a common fermion. From quantum information insight, the topological qubit is EPR-like, because it encode quantum state non-locally. Both characteristics explain its resistance to local disturbances. Given that topological excitations are robust against local perturbations, The question that may arise is whether topological qubits are also robust against decoherence, when they are considered as an open quantum system coupling to the non-topological environment. The open system setup is more logical and realistic  when performing quantum computations. Since quantum information is carried by physical excitations, stability against decoherence  indicates stability against local perturbations, but the opposite is not true. Even if local excitations are stable to local perturbations, however the quantum information carried by the topological qubit may still penetrate into the environment. However, since topological qubit is  non-local, its interaction with the environment is quite different from that of conventional fermions, and quantum quantum decoherence behaviors are expected to be unusual. This
motivates us, in this work, to examine QSL time for decoherence of topological qubit. 

In this paper the QSL time for the dynamics of a topological qubit realized by two Majorana modes coupled to a fermionic and bosonic Ohmic-like reservoir is discussed in detail.
\section{Decoherence of topological qubits}
\begin{figure}[ht] 
\centering
\includegraphics[width=8cm]{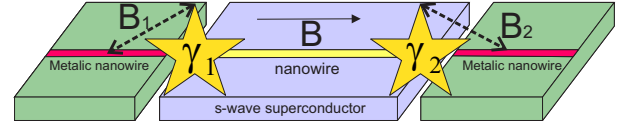}
\caption{Schematic representation of topological qubit
realized by two Majorana modes $\gamma_1$ and $\gamma_2$, interacting with a fermionic environment.}\label{fig1}
\end{figure}

A topological qubit composed of two Majorana modes of the one-dimensional Kitaev’s chain that are spatially separated. These Majorana modes are located at the two ends of a quantum wire and are denoted by $\gamma_1$ and $\gamma_2$, and the following relations are established for them
\begin{equation}\label{gamma}
\gamma_{a}^{\dagger}=\gamma_{a}, \quad\left\{\gamma_{a}, \gamma_{b}\right\}=2 \delta_{a b},
\end{equation}
where $a, b \in\{1,2\}$. The Majorana modes are influenced by their surroundings in an incoherent way, which causes the decoherence of the topological qubit. The Hamiltonian describing the intended general system is defined as follows
\begin{equation}
\hat{H}=\hat{H}_S+\hat{H}_E+\hat{H}_I,
\end{equation}
where $\hat{H}_S$ represents the Hamiltonian of topological qubit system, $\hat{H}_E$ describes the Hamiltonian of the environment and $\hat{H}_I$ is the interaction Hamiltonian describes the influence of the environment on topological qubit which reads
\begin{equation}
\hat{H}_I=G_1 \gamma_1 \hat{\mathcal{Q}}_1 + G_2 \gamma_2 \hat{\mathcal{Q}}_2,
\end{equation}
where $G_{1(2)}$ describes the real coupling constant and $\hat{Q}_{1)(2)}$ is the composite operator consist of the electron creation operator $a^{\dag}$   and annihilation operator $a$ . According to hermeticity condition of interaction Hamiltonian i.e. $\hat{H}_I^{\dag}=H_I$ we have
\begin{equation}
\hat{\mathcal{Q}}_a^{\dag}=-\hat{\mathcal{Q}}_a.
\end{equation} 
When the system is affected by a fermionic environment, the Majorana modes are located at two ends of a quantum wire which is placed over a s-wave superconductor. Majorana modes are affected by a magnetic field $B$ whose direction is along the quantum wire. Each of the Majorana modes is paired with a metallic nanowire through  a tunnel junction with tunneling strength $B_i$ that is adjusted by an external gate voltage. The schematic of this type of interaction is drawn in Fig. (\ref{fig1}).

\begin{figure}[ht] 
\centering
\includegraphics[width=8cm]{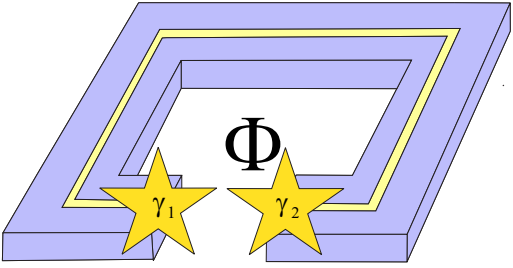}
\caption{Schematic representation of topological qubit
realized by two Majorana modes $\gamma_1$ and $\gamma_2$, interacting with a bosonic environment.}\label{fig2}
\end{figure}

When the system is affected by a bosonic environment, the Majorana modes are placed at two ends of a quantum ring with an empty space between them. In  this case the some
environmental bosonic operator interact with the two Majorana modes locally. the frequency dependence in the bosonic environment is provided by a magnetic flux $\Phi$ passing through the quantum ring. The schematic of this type of interaction is drawn in Fig. (\ref{fig2}).
In this work we consider Ohmic-like environmental spectral density $J(\omega) \propto \omega ^s$, for both fermionic and bosonic environment. When $s<1$, $s=1$ and $s>1$ we have sub-Ohmic, Ohmic and super-Ohmic environment respectively.

Before interaction, the Majorana modes form a topological qubit with states $\vert 0 \rangle
$ and $\vert 1 \rangle$, which are related as follows 
\begin{equation}
\frac{1}{2}\left(\gamma_{1}-i \gamma_{2}\right)|0\rangle=|1\rangle, \quad \frac{1}{2}\left(\gamma_{1}+i \gamma_{2}\right)|1\rangle=|0\rangle
\end{equation}
Given that $\gamma_a$'s must satisfy Eq. (\ref{gamma}), they can be selected as
\begin{equation}
\gamma_{1}=\sigma_{1}, \quad \gamma_{2}=\sigma_{2}, \quad i \gamma_{1} \gamma_{2}=\sigma_{3},
\end{equation}
where $\sigma_i$'s are Pauli matrices.

Here it is assumed that the state of the whole system $S+E$ is product i.e. $\rho_S(0) \otimes \rho_E$ and $\rho_{S}(0)=\sum_{i, j=0}^{1} \rho_{i j}|i\rangle\langle j|$ is the initial state of the topological qubit.  For the case where the topological qubit is affected by a fermionic environment, the state of the topological qubit at time $t$ is obtained as
\begin{equation}
\rho_{S}^{F}(t)=\frac{1}{2}\left(\begin{array}{cc}
1+\left(2 \rho_{00}-1\right) \alpha^{2}(t) & 2 \rho_{01} \alpha(t) \\
2 \rho_{10} \alpha(t) & 1+\left(2 \rho_{11}-1\right) \alpha^{2}(t)
\end{array}\right),
\end{equation}
and for the case where the topological qubit is influenced by a bosonic environment, the state of the topological qubit at time $t$ is obtained as 
\begin{equation}
\rho_{S}^{B}(t)=\left(\begin{array}{cc}
\rho_{00} & \rho_{01} \alpha(t) \\
\rho_{10} \alpha(t) & \rho_{11}
\end{array}\right),
\end{equation}
where 
\begin{equation}\label{de}
\alpha(t)=e^{-2 B^{2}\left|\beta_{F, B}\right| \mathcal{I}_{s}(t)},
\end{equation}
and decay parameter $\mathcal{I}_{s}(t)$ is given by
\begin{equation}
\mathcal{I}_{s}(t)=\left\{\begin{array}{ll}
2 \Gamma_{0}^{s-1} \Gamma\left(\frac{s-1}{2}\right)\left(1-{ }_{1} F_{1}\left(\frac{s-1}{2} ; \frac{1}{2} ; \frac{-\Gamma_{0}^{2} t^{2}}{4}\right)\right) & s \neq 1 \\
\frac{\Gamma_{0}^{2} t^{2}}{2}{ }_{2} F_{2}\left(\{1,1\} ;\left\{\frac{3}{2}, 2\right\} ; \frac{-\Gamma_{0}^{2} t^{2}}{4}\right) & \quad s=1
\end{array}\right.
\end{equation}
where $\Gamma_0$ describes the cutoff frequency of the environment, $\Gamma(z)$ is the Gamma function and ${ }_{i} F_{j}$ is the generalized hypergeometric function. $\beta_F$ and $\beta_B$  are time-independent coefficients in Fermion and boson environments, respectively and are defined as 
\begin{equation}
\beta_{F}=\frac{-4 \pi}{\Gamma\left(\frac{s+1}{0}\right)}\left(\Gamma_{0}\right)^{-(s+1)}
\end{equation}
and
\begin{equation}
\beta_{B}=\left\{\begin{array}{ll}
-\frac{N_{s s}^{2} \Gamma(3-\Delta) \epsilon^{2(\Delta-4)}}{4 \pi^{2} \Gamma(\Delta-2) 2^{2} \Delta-5} \sin \pi \Delta & 2<\Delta \notin \mathbf{N} \\
-\frac{N_{s c}^{2} e^{2(\Delta-4)}}{4 \pi(\Delta-3) ! 22^{2 \Delta-5}} & 2 \leq \Delta \in \mathbf{N}
\end{array}\right.
\end{equation}
where $N_{sc}$ represents the number of degrees of freedom of the dual conformal field theory, $\epsilon$ is the $UV$ cutoff of the length, $\Delta=(s+4)/2$ shows the conformal dimension and $N$ stands for the set of natural numbers. 
\section{Quantum speed limit time}
Quantum mechanics sets a bound on the evolution speed of a quantum process for a closed or open quantum systems. In general, the minimum time required to transform from an initial state to a target state is known as QSL time. Mandelstam and Tamm have introduced the QSL time bound known as (MT) bound for close quantum system as \cite{11} 
\begin{equation}\label{m1}
\tau \geq \tau_{Q S L}=\frac{\pi \hbar}{2 \Delta \mathcal{E}},
\end{equation}  
where $\Delta \mathcal{E} =\sqrt{ \langle \hat{H}^2 \rangle  - \langle \hat{H}\rangle^2}  $ is the variance of energy of initial state and $\hat{H}$ is the time-independent Hamiltonian describing the evolution of the closed quantum system. Another bound for closed quantum systems, have introduced by Margolus and Levitin \cite{12}. It is known as (ML) bound, which is defined as follows  
\begin{equation}
\tau \geq \tau_{Q S L}=\frac{\pi \hbar}{2 \mathcal{E}},
\end{equation}
where $\mathcal{E}= \langle \hat{H} \rangle$. By combining the two ML and MT bounds for close quantum system, one can obtain an unified bound  for QSL time as follows
\begin{equation}
\tau \geq \tau_{Q S L}=\max \left\{\frac{\pi \hbar}{2 \Delta E}, \frac{\pi \hbar}{2 E}\right\}.
\end{equation}
In the real world, the interaction of a system with its environment is inevitable, so in practice the study of open quantum systems is of particular importance. The evolution of an open quantum system is described using a time-dependent master equation as 
\begin{equation}
\dot{\rho}_t=\mathcal{L}_t \rho_t,
\end{equation}
where $\rho_t$ is the state of the open quantum system at time $t$ and $\mathcal{L}_t$ is the positive generator \cite{15}.The main goal here is to find the minimum time required to evolve from an initial state $\rho_\tau$ to a target state $\rho_{\tau+\tau_D}$ for open quantum system, where $\tau$ is initial time and $\tau_D$ is driving time. This minimal time is called QSL time. In order to quantify the bound of QSL time one should use an appropriate distance measure. In Refs. \cite{23,28}, the authors use relative purity to quantify the bound for QSL time. A notable feature of the bound defined by them is that their bound can also be used for mixed initial states. The relative purity between the initial state $\rho_\tau$ and the target state $\rho_{\tau+\tau_D}$ is defined as follows
\begin{equation}
f\left(\tau+\tau_{D}\right)=\frac{\operatorname{tr}\left(\rho_{\tau} \rho_{\tau+\tau_{D}}\right)}{\operatorname{tr}\left(\rho_{\tau}^{2}\right)}
\end{equation}
By following the method given in Ref.\cite{28}, the
ML bound of QSL-time can be obtain as 
\begin{equation}
\tau \geq \frac{\left|f\left(\tau+\tau_{D}\right)-1\right| t r\left(\rho_{\tau}^{2}\right)}{\sum_{i=1}^{n} \kappa_{i} \varrho_{i}}
\end{equation}
where $\kappa$ and $\varrho$ are the singular value of $\mathcal{\rho_t}$ and $\rho_\tau$ respectively and $\overline{\square}=\frac{1}{\tau_{D}} \int_{\tau}^{\tau+\tau_{D}} \square d t$.  By following a similar method, the MT
bound of QSL-time for open quantum system can be obtain as 
\begin{equation}
\tau \geq \frac{\left|f\left(\tau+\tau_{D}\right)-1\right| \operatorname{tr}\left(\rho_{\tau}^{2}\right)}{\sqrt{\sum_{i=1}^{n} \kappa_{i}^{2}}}
\end{equation}
By combining ML and MT bounds, an unified bound can be achieved as follows
\begin{equation}
\begin{aligned}
\tau_{(Q S L)}=& \max \left\{\frac{1}{\sum_{i=1}^{n} \kappa_{i} \rho_{i}}, \frac{1}{\sqrt{\sum_{i=1}^{n} \kappa_{i}^{2}}}\right\} \\
& \times\left|f\left(\tau+\tau_{D}\right)-1\right| t r\left(\rho_{\tau}^{2}\right).
\end{aligned}
\end{equation}
\section{Results}
In this section we want to find the QSL time for the topological qubit when they interact with the bosonic or fermionic environment. Let us consider the initial mixed state for topological qubit as 
\begin{equation}\label{initial}
\rho_{0}=\frac{1}{2}\left(\begin{array}{cc}
1+v_{z} & v_{x}-i v_{y} \\
v_{x}+i v_{y} & 1-v_{z}
\end{array}\right).
\end{equation}
When topological qubit interacts with fermionic environment its time evolution reads
\begin{equation}
\rho_S^F(t)=\frac{1}{2}\left(\begin{array}{cc}
1+v_{z}^{\prime} & v_{x}^{\prime}-i v_{y}^{\prime} \\
v_{x}^{\prime}+i v_{y}^{\prime} & 1-v_{z}^{\prime}
\end{array}\right). 
\end{equation}
Now we can find the singular value of $\rho_\tau$ and $\mathcal{L}_t(\rho_t)$. the singular value of $\rho_\tau$ are
\begin{equation}
\begin{array}{l}
\varrho_{1}=\frac{1}{2}\left(1-\sqrt{v_{x}^{\prime 2}+v_{y}^{\prime 2}+v_{z}^{\prime 2}}\right) \\
\varrho_{2}=\frac{1}{2}\left(1+\sqrt{v_{x}^{\prime 2}+v_{y}^{\prime 2}+v_{z}^{\prime 2}}\right)
\end{array}
\end{equation}

where $v_{x}^{\prime}=\alpha(\tau) v_x$, $v_{y}^{\prime}=\alpha(\tau) v_y$ and  $v_{z}^{\prime}=\alpha(\tau)^{2} v_{z}$. The singular value $\kappa_i$ of $\mathcal{L}_t(\rho_t)$ can be written as 
\begin{equation}
\begin{array}{l}
\kappa_{1}=\kappa_{2}=\frac{1}{2} \vert \dot{\alpha}(t)\sqrt{v_x^2+v_y^2+4 \alpha (t)^2 v_z^2} \vert ,
\end{array}
\end{equation}
Therefore, it can be concluded that $\varrho_1 \kappa_1 + \varrho_2 \kappa_2$ is always less than $\sqrt{\kappa_1^2+\kappa_2^2}$ and so the ML bound on QSL time is tighter than MT bound for open quantum systems.  For the bosonic environment we follow our calculations in the same way as before . We consider the initial state in Eq.(\ref{initial}). For the case in which the topological qubit is influenced by bosonic environment its time evolution reads
\begin{equation}
\rho_S^B(t)=\frac{1}{2}\left(\begin{array}{cc}
1+v_{z}^{\prime} & v_{x}^{\prime}-i v_{y}^{\prime} \\
v_{x}^{\prime}+i v_{y}^{\prime} & 1-v_{z}^{\prime}
\end{array}\right). 
\end{equation}
The singular value of $\rho_\tau$ and $\mathcal{L}_t(\rho_t)$ can be obtain easily like fermionic environment. The singular value of $\rho_\tau$ are
\begin{equation}
\begin{array}{l}
\varrho_{1}=\frac{1}{2}\left(1-\sqrt{v_{x}^{\prime 2}+v_{y}^{\prime 2}+v_{z}^{\prime 2}}\right) \\
\varrho_{2}=\frac{1}{2}\left(1+\sqrt{v_{x}^{\prime 2}+v_{y}^{\prime 2}+v_{z}^{\prime 2}}\right)
\end{array}
\end{equation}

where $v_{x}^{\prime}=\alpha(\tau) v_x$, $v_{y}^{\prime}=\alpha(\tau) v_y$ and  $v_{z}^{\prime}= v_{z}$. The singular value $\kappa_i$ of $\mathcal{L}_t(\rho_t)$ can be written as 
\begin{equation}
\begin{array}{l}
\kappa_{1}=\kappa_{2}=\frac{1}{2} \vert \dot{\alpha}(t)\sqrt{v_x^2+v_y^2} \vert ,
\end{array}
\end{equation}

In this work we consider the maximally coherent initial state with $v_x=v_y=1/\sqrt{2}$ and $v_z=0$ . So the QSL time for topological qubit inside both fermionic and bosonic environment can be obtained as 
\begin{equation}
\tau_{QSL}=\frac{\vert \alpha(\tau)^2-\alpha(\tau)\alpha
(\tau+\tau_D)\vert}{\frac{1}{\tau_D}\int_\tau^{\tau+\tau_D}\vert \dot{\alpha}(t)\vert}.
\end{equation}
Although we have reached a similar relation for two different environments, however due to the difference in $\alpha(t)$ coefficient for both bosonic and fermionic environments, the results will definitely be different. 
\begin{figure}[ht] 
\centering
\includegraphics[width=8cm]{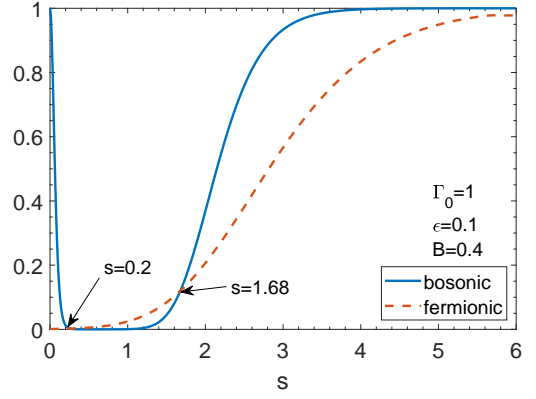}
\caption{(Color online) QSL time versus Ohmic parameter $s$ for different bosonic and fermionic environment, B=0.4, $\tau_D=1$ and $\tau=1$.}\label{fig3}
\end{figure}
\begin{figure}[ht] 
\centering
\includegraphics[width=8cm]{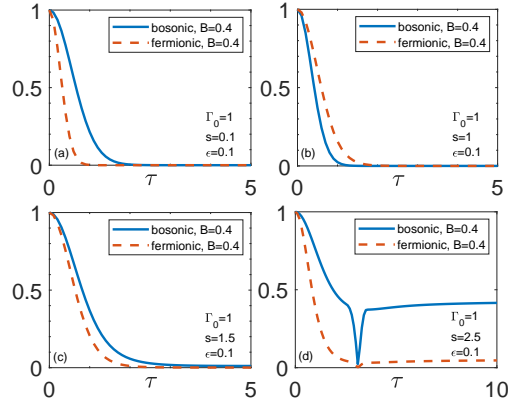}
\caption{(Color online) QSL time versus initial time parameter $\tau$ for different bosonic and fermionic environment when $\tau_D=1$ (a)s=0.1, (b)s=1, (c)s=1.5 and (d)s=2.5.}\label{fig4}
\end{figure}

\begin{figure}[ht] 
\centering
\includegraphics[width=8cm]{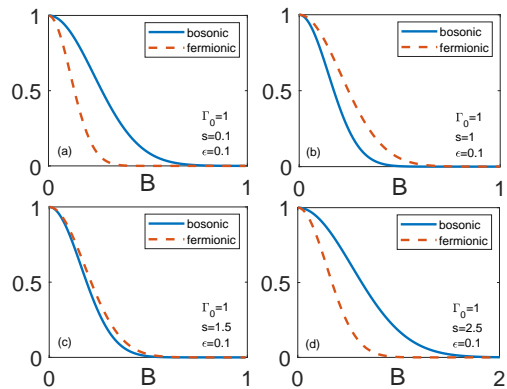}
\caption{(Color online) QSL time versus magnetic field parameter $B$ for different bosonic and fermionic environment when $\tau_D=1$ (a)s=0.1, (b)s=1, (c)s=1.5 and (d)s=2.5}\label{fig5}
\end{figure}
In Fig.\ref{fig3}, QSL time is plotted as a function of Ohmic parameter for both bosonic and fermionic environment. As can be seen for bosonic environment the QSL time reaches to zero for small value of $s$ and increases again with increasing $s$. The situation is slightly different for the fermion environment. For this environment, the QSL time starts from zero for small values of $s$ and increases steadily with increasing $s$. It is important to note that in any case the QSL time is smaller than the driving time $\tau_D$ otherwise the miscalculations are done. Also, as can be seen from the diagram, for the interval $s<0.2$ and $s \in [0.2,1.68]$, the QSL time in the fermionic environment is greater than the bosonic environment, and in other intervals the QSL time in the bosonic environment is higher than QSL time in the fermionic environment. 

In Fig.\ref{fig4}, the QSL time is plotted in terms of initial time $\tau$ for different value of Ohmic parameter for both fermionic and bosonic environment. Fig.\ref{fig4}(a) shows the QSL time as a function of initial time $\tau$ for sub-Ohmic environment. From Fig.\ref{fig4}(a) one can see for sub-Ohmic bosonic and fermionic environment the QSL time starts to decrease at the beginning of the evolution until it reaches to zero. It is also observed that the QSL time for bosonic environment is greater than fermionic environment  since the value of $s$ is smaller than $0.2$. In Fig.4(b), QSL time is plotted for Ohmic bosonic and fermionic environment. For Ohmic environment the behavior of QSL time is similar with sub-Ohmic environment. It is also observed that the QSL time for fermionic environment is greater than QSL time for bodsonic environment. In Fig.\ref{fig4}(c) and Fig.\ref{fig4}(d), the QSL time is plotted for both super-Ohmic bosonic and fermionic environment. As can be seen by increasing $s$ the QSL time increases for both fermionic and bosonic environment. As can be seen from Fig.\ref{fig4}(d) for super-Ohmic environment with larger value of $s$ due to the occurrence of coherence trapping \cite{Addis}, the QSL time  would be gradually trapped to a fixed value, and so leads to a uniform evolution speed for the open system.

In Fig.\ref{fig5}, the QSL time is plotted as a function of magnetic parameter $B$ for both bosonic and fermionic environment. As can be seen from Fig.\ref{fig5} the QSL time decreases with increasing magnetic parameter for sub-Ohmic, Ohmic and super-Ohmic bosonic and fermionic environment. 




\section{Conclusion}
In this work, we have studied the QSL time for topological qubit influenced by fermionic and bosonic environment. We assume that the environments have an Ohmic-like spectral density. We have shown that for Ohmic and sub-Ohmic environment the QSL time reduces with the starting point in time, in other words, it can be said that the open system experiences a speeded-up dynamics evolution process. It is observed that  for super-Ohmic environment with increasing Ohmic parameter the quantum speed limit time gradually reaches to a constant value and thus the speed of evolution reaches to a uniform value. We have also studied the effect of magnetic field on QSL time. It has shown that the QSL time decreases with increasing the value of magnetic field for both bosonic and fermionic environment with different Ohmic parameter.

\end{document}